# TRAPPING OF COMPENSATING CHARGES IN CORONA POLED PVDF FILMS

A. F. Butenko, S. N. Fedosov, A. E. Sergeeva
*Department of Physics, Odessa National Academy of Food Technologies, Ukraine*

Experimental evidence is provided on close relation between polarization and space charge in corona poled PVDF films. It is shown that the depolarizing field is compensated by charges trapped in macroscopic transition zones, by which polarized parts of the volume are separated from non-polarized ones. The compensating charges are either injected in the bulk, or created inside due to the thermal emission and then trapped when the ferroelectric polarization in crystallites is formed.

## 1. Introduction

PVDF and some of its copolymers belong to a class of ferroelectric polymers, because they possess a spontaneous switchable polarization in one of the crystalline polymorphs called *b* phase [1, 2]. It has been proved experimentally that the piezoelectric and pyroelectric coefficients of PVDF are proportional to the magnitude of the remnant polarization [3]. At the same time, there is strong evidence that injected space charge also affects specific properties of ferroelectric polymers [4-10]. The role of the space charge has been discussed for the past twenty years or so, but no agreement has been reached on this point so far. Some researchers believe that presence of the space charge in PVDF is important, while others think its influence is negligible [11]. It is implied in many cases that polarization and space charge do not depend on each other, although, in our opinion, correlation between them must be considered.

It is known that switching of polarization in ferroelectrics is accompanied by appearance of the depolarizing field that tries to restore the original electric state with zero average polarization [12]. If this field is not neutralized (or compensated), the residual polarization is low and unstable. The compensation of the depolarizing field in conventional ferroelectrics is usually done either by charges adsorbed on the surfaces, or due to redistribution of charges on electrodes. In the case of ferroelectric polymers the phenomenon is more complicated, because polarized crystallites are dispersed in a disordered amorphous phase. It is not known where the compensating charge is trapped [13-15] and how the charge trapping can be recognized and identified experimentally.

It has been found that the poled PVDF is less conductive than the unpoled one [5, 16]. Measurement of the spatial distribution of polarization has shown that injection of charges takes place disturbing the uniformity of polarization [6,7,9,13,14]. We proposed a model for the buildup of polarization in PVDF and other ferroelectric polymers assuming that trapped charges and polarization form a self-consistent and well balanced system [8,17-19]. In this work we provide more experimental data on trapping of compensating charges in corona poled PVDF films. It is shown by appropriate calculations that the trapped charges reside at the boundaries of the macroscopic polarized zones.

## 2. Experimental

The study was performed on 25 μm-thick uniaxially stretched PVDF films containing amorphous and crystalline phases in almost equal proportion. The crystallites consisted of a ferroelectric *b* phase (56%) and a non polar *a* phase (44%). Samples were electroded from one side by evaporation of aluminum and placed in a corona triode with a non-metallized surface exposed to corona ions. The corona discharge was initiated by a pointed tungsten electrode kept at a constant potential of -20 kV. The grid was made vibrating in order to measure effective surface (electret) potential by Kelvin’s method directly during poling and immediately after its completion [8].

Three different regimes of poling were applied. In the first series of experiments, the quasi-stationary *I-V* characteristics have been measured at room temperature by increasing a negative dc voltage at the control grid in steps of 100 V from 0.6 to 3.0 kV and recording the poling current. The samples were kept at each voltage for 15 min before application of the next voltage step. The experiment was repeated again on the already poled samples.

In the second series, a constant current of 80 μA/m$^2$ was applied periodically for 5 min at room temperature followed by intervals of the same duration when poling was discontinued [20]. In this case, the grid was kept at a constant potential of -3 kV, while the voltage at the corona electrode was controlled automatically by a feedback circuit, so that complete poling was performed in 11 charging-discharging cycles. The growth and the decay of the electret potential were recorded during the corresponding fractions of each cycle.

In the third series of experiments, temperature increased linearly from 20 to 100 °C at the rate of 3 K/min, while the grid voltage, and consequently the electret potential, were kept constant. Then the samples were cooled down to room temperature under the applied field and the experiment was repeated again under the same conditions. Describing experimental results we use a concept of the apparent conductivity $g$ defined as $g=Ix_o/VA$, where $I$ is the measured current, $V$ the applied voltage, $x_o$ the thickness of the sample, $A$ the surface area.

## 3. Results and discussion

There are three sections of the *I-V* characteristics corresponded to three stages of initial poling (Fig. 1). Initially current grows faster than the applied voltage until the apparent conductivity reaches its maximum at the poling field of 55 MV/m, being very close to the coercive field of PVDF [21]. At the second stage, the conductivity drops abruptly to the minimum at the field of 110 MV/m, followed by a slight increase at higher voltages. Repeated measurements on the same sample show that low values of the apparent conductivity are retained in the poled samples indicating that the changes occurred during initial poling are irreversible. The field at the second stage was higher than the coercive value, therefore the main part of the ferroelectric polarization has been formed during this stage. Since the decrease of the apparent conductivity and the growth of polarization occur in the same range of the poling field, it is very probable that these two phenomena are interrelated.

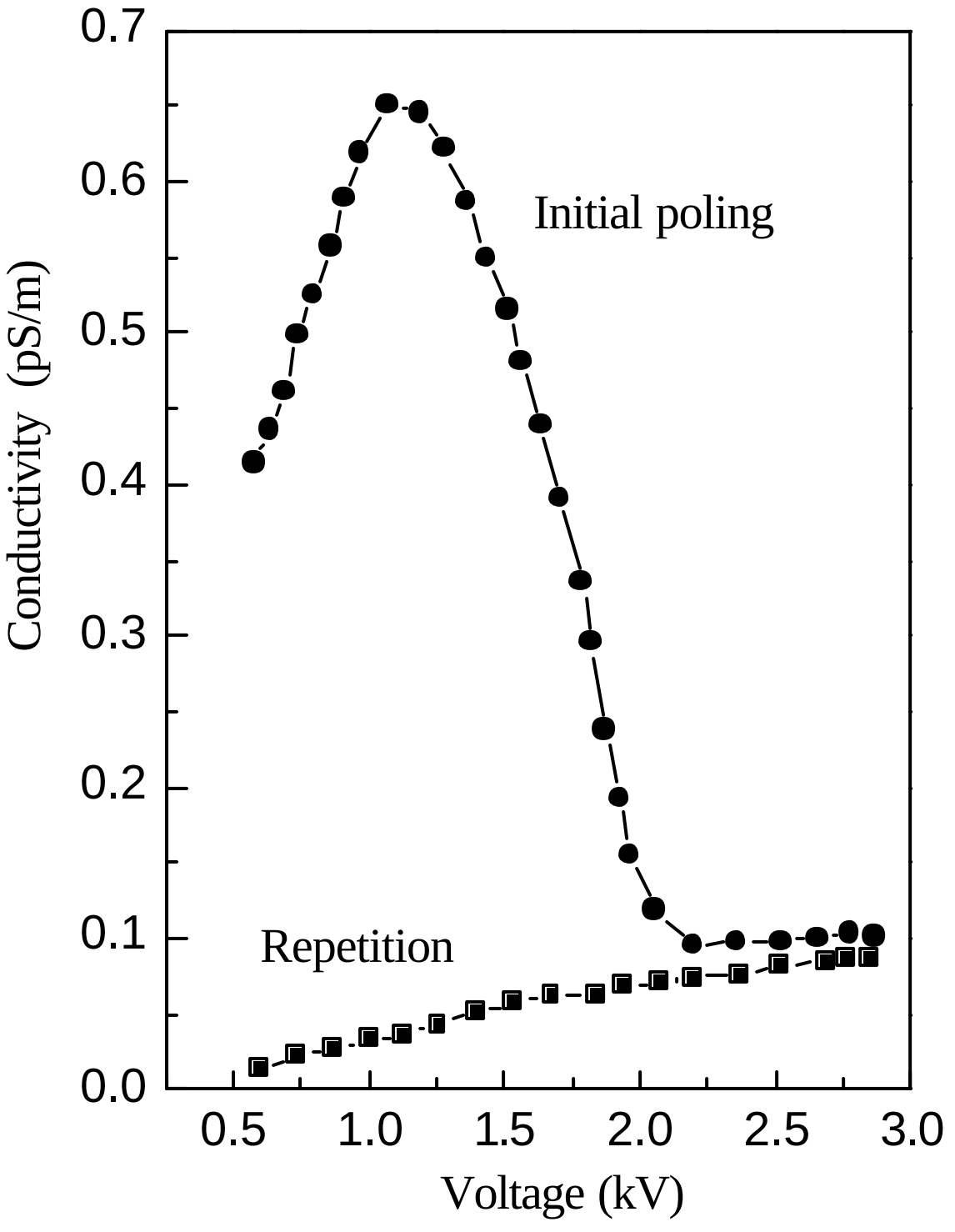

Fig.1. Dependence of the apparent conductivity on applied voltage during initial poling and subsequent repetition.

Poling conditions have been changed in the second series of experiments, so that the constant current was applied periodically. It is seen in Fig. 2 that the electret potential after completion of the first poling cycle decays almost to zero in 300 s with the time constant of about 200 s corresponded to Maxwell's relaxation time for PVDF at room temperature [22]. At the same time, the electret potential becomes more stable if the number of the poling cycles increases. For example, the relaxation time of 1300 s was observed after application of 11 poling cycles showing that the apparent conductivity during poling became 6.5 times smaller than its initial value.

At the third series, we observed anomalous behavior of the poling current during the linear increase of temperature at a constant applied voltage. As one can see in Fig. 3, the apparent conductivity decreased or remained constant during poling depending on the value of applied voltage, but the activation energy of the apparent conductivity did not change, being about 0.8 eV in unpoled and poled samples, as follows from the slope of graphs in Fig. 3. It is assumed here that the temperature dependence of the conductivity obeys the Arrenius law

$$g = emn_o \exp\left(-\frac{Q}{kT}\right) \quad (1)$$

where $Q$ is the activation energy, $k$ is Boltzmann's constant, $T$ the absolute temperature, $e$ the elementary charge, $m$ the mobility of the charge carriers. It seems that the coefficient $n_o$ becomes smaller during poling, as if free charge carriers are deeply trapped somewhere.

Although poling conditions were different in three series of the experiments, the obtained results had two common features. First, the apparent conductivity decreased abruptly when the ferroelectric polarization started to grow. Second, the decrease in conductivity was irreversible independently on the poling conditions.

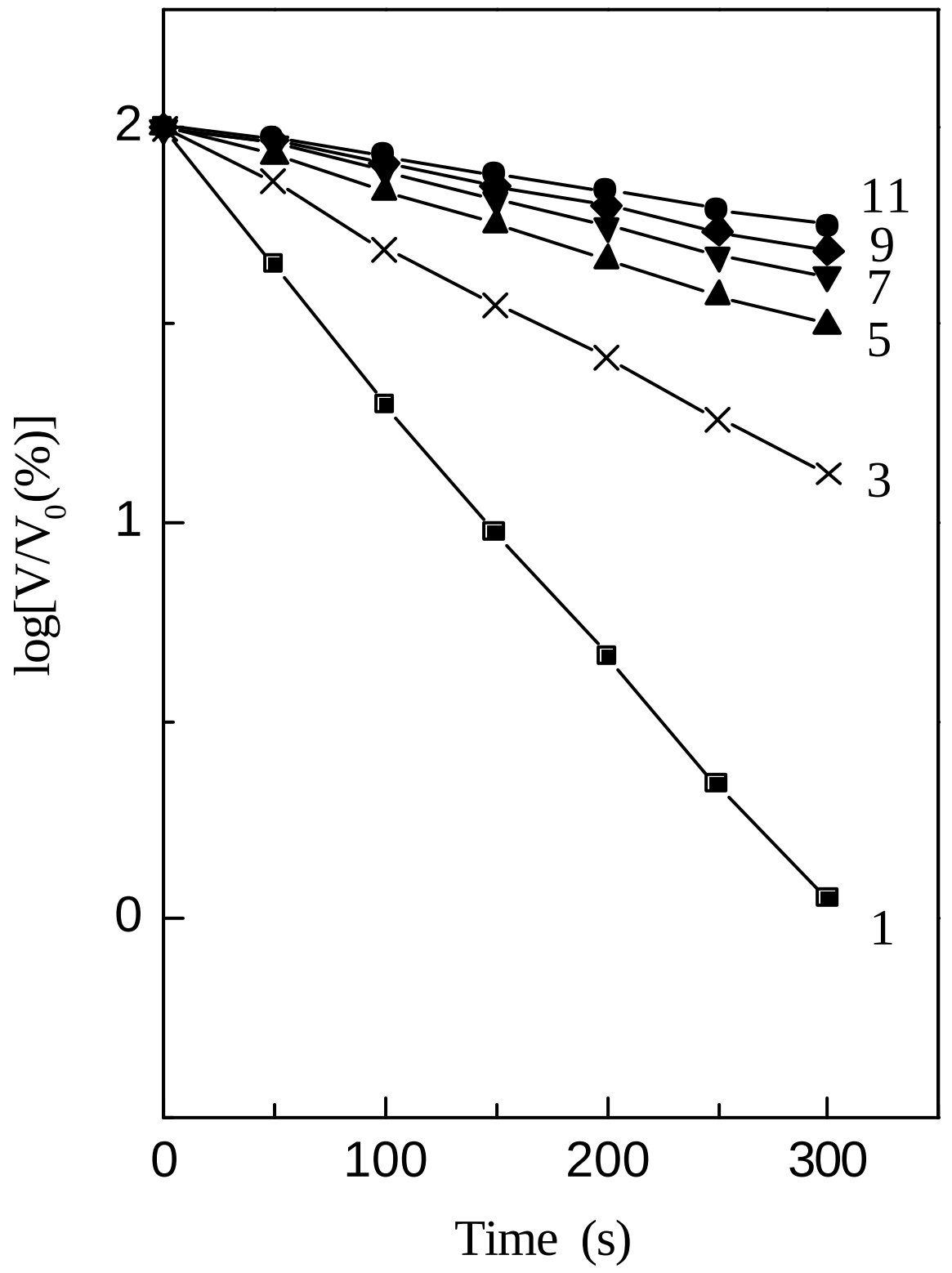

Fig.2. Decay of the electret potential in intervals between periodic applications of the constant poling current. Numbers correspond to the poling cycles.

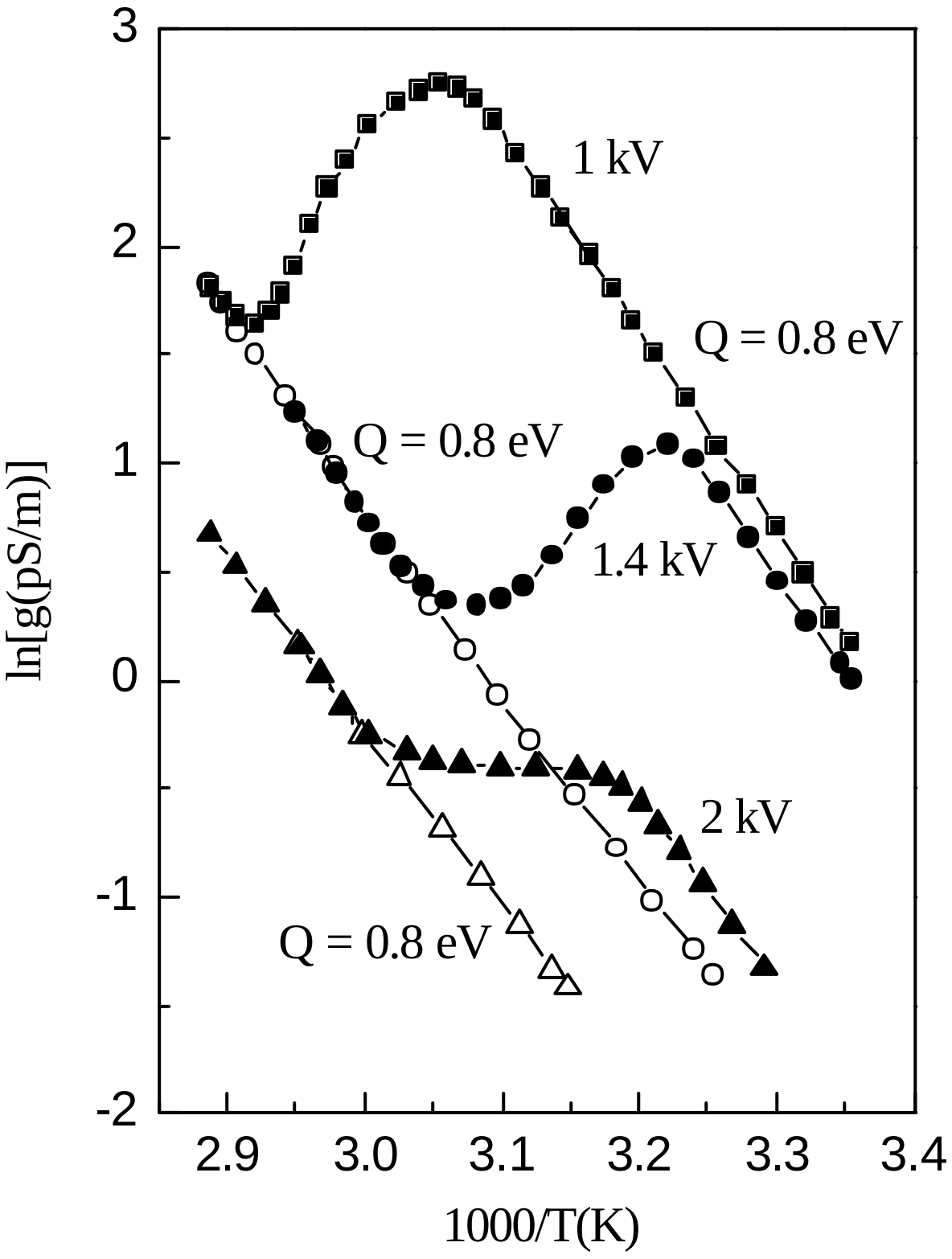

Fig.3. Dynamics of the apparent conductivity during initial thermally stimulated poling (solid symbols) and during repeated charging of poled samples (hollow symbols). Poling voltages and activation energies are also shown.

It is known that the total poling current $i(t)$ has three components, namely the capacitive, the polarization and the conductive ones according to the following equation

$$i(t) = e_o e \frac{dE}{dt} + \frac{dP}{dt} + gE \quad (2)$$

where $i(t)$ is the current density, $e_o$ is the permittivity of vacuum, $e$ the dielectric constant, $E$ the field, $P$ the ferroelectric polarization. The capacitive component represented by the first term at the right hand side of Eq. (2) can be neglected, because we applied constant field in the first and the third series of experiments, while the voltage (electret potential) changed very slowly in the second series. It is not possible to calculate exactly the contribution of the polarization component to the total current, but it can be estimated by comparing the integral of the total current with the highest theoretically obtainable value of the residual polarization in PVDF.

Our calculations show that the integral under the current peak during the thermally stimulated poling at 1 kV is equal to 260 mC/m$^2$ , while the highest possible magnitude of polarization in PVDF is only 65 mC/m$^2$ [2, 3]. Therefore, the polarization component amounts about 25% of the total current. Similar results were obtained from integration of absorption currents (not shown here) during the first series of experiments indicating that the main contribution to the measured current gives the conductive component.

It is obvious that the apparent conductivity $g$ depends on the density of free carriers composed of intrinsic carriers $n_o$ and injected ones $n_{i.}$ Considering one kind of mobile carriers, we can write

$$g = em(n_o + n_i) \tag{3}$$

While $n_o$ depends mostly on temperature, $n_i$ is affected by poling conditions. It seems that charges in our experiments were injected in the bulk during the first stage of poling and then trapped when the ferroelectric polarization started to grow. We have shown that favorable conditions for charge trapping exist at the boundaries of polarized crystallites [15, 19]. Therefore, it is possible to calculate what would be the density of the compensating charges, if they are trapped at the crystallite boundaries. Assuming that PVDF has a brick-like structure [23], we get

$$n'_c = ab\frac{P_s}{eV_c} \tag{4}$$

where $a$=7.5 nm and $b$=3.5 nm are average length and width of a crystallite [23], $V_c$ =1.2x10$^{-25}$ m$^3$ is the volume occupied by one crystallite together with the attached amorphous phase, $P_s$=0.14 C/m$^2$ [2, 12] is the spontaneous polarization in crystallites. Calculation gives the value of $n'_c$=2·10$^{20}$ cm$^{-3}$, while the real density of free charge carriers available for trapping is only $n_c$ =10$^{12}$–10$^{14}$ cm$^{-3}$, if typical for PVDF values of the apparent conductivity (10$^{-10}$–10$^{-12}$ S/m) and the mobility of the charge carriers (2·10$^{-8}$ cm$^2$/Vs) are assumed [24].

The difference in value of $n_c$ and $n'_c$ is so enormous that even a massive injection of charges will not provide sufficient number of carriers for trapping at the boundaries of all crystallites. Therefore, compensating charges are not trapped there, although energetic conditions are favorable for this. How then the depolarizing field in crystallites is compensated? To answer the question, one should consider Poisson's equation

$$e_o e\frac{dE}{dx} = r(x) + \frac{dP}{dx} \tag{5}$$

where $r(x)$ is the volume density of the space charge, $x$ the coordinate in the thickness direction. When poling is completed and the sample is short-circuited, the field everywhere in the bulk becomes zero due to redistribution of the mobile charges. The characteristic time for this process is of the order of Maxwell's relaxation time. Then one gets from Eq. (5) at $E$=0

$$r(x) = -\frac{dP}{dx} \qquad (6)$$

indicating that the charges can reside in transition zones where polarization is non-uniform and $dP/dx \neq 0$. Thus, conditions for charge trapping are favorable not only at the boundaries of crystallites, but also in macroscopic transition zones separating polarized part of the volume from non-polarized one, as it is shown schematically in Fig. 4.

Trapping of charges in transition zones is very important, because the charges indeed compensate the depolarizing field in the crystallites. To prove this, we consider, for example, the case of the uniform distribution of polarization in the bulk with narrow transition zones near surfaces [9]. Under equilibrium conditions, the density $s_r$ of the real charge at the surfaces is equal to the density of the polarization charge $s_p$ and to the magnitude of the average polarization *P¢*, i.e. $s_r = s_p =$ *P¢*. In its turn, *P¢* $= x_c P_s$, where $x_c$ is the crystallinity. The field applied to a dipole inside a crystallite is zero, because the number of positively charged layers are equal to that of the negatively charged ones both above this dipole and below it. Therefore, the depolarizing field inside the crystallites is completely compensated by the field of charges trapped either on the surface, or in transition zones. It is easy to show that in both cases the charge density related to the unit area does not depend on thickness of the transition zone and equal to the maximum value of polarization. For example, in order to compensate the depolarizing field in polarized crystallites, the volume density $n_v$ of the trapped charge should correspond to the following equation

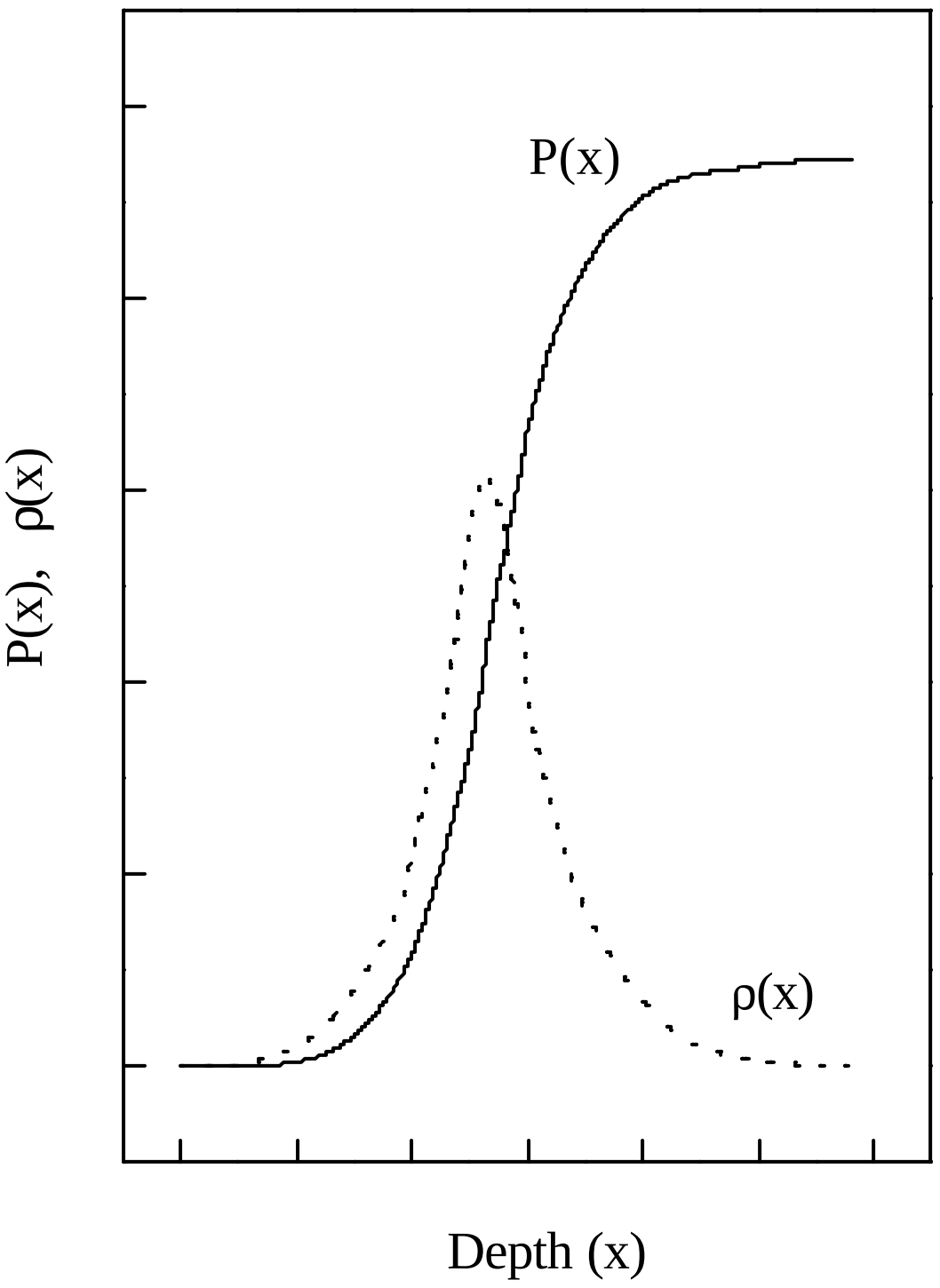

Fig.4. Typical distribution of polarization P(x) and space charge ρ(x) in the thickness direction.

$$s_r = x_c P_s = e n_n x_o \qquad (7)$$

Assuming $x_o$=20 µm, $x_c$=0.5, $P_s$ =0.14 C/m$^2$ one gets $n_v$ =5·10$^{16}$ cm$^{-3}$ which is lower than the density of the intrinsic carriers $n_c$ . This result is rather important, because it shows that there must be a sufficient source of free charges during poling, so that they can be trapped in transition zones in order to compensate the depolarizing field in the crystallites. The charges can be either injected, as in the case of corona poling, or emitted thermally inside the sample. For example, the required charge can be injected in 15 min, if poling is performed at room temperature with a constant current of 160 µA/m$^2$.

It is known that PVDF poled at elevated temperatures has higher value of the residual polarization than that poled at room temperature. The thermal activation, most probably, not only improves alignment of dipoles, as in the case of conventional polar electrets, but also provides the high density of free charge carriers necessary for further compensation of the depolarizing field. The electret component of polarization in PVDF is thermodynamically unstable and decays with time, negatively effecting the stability of piezoelectric and pyroelectric coefficients.

The field in amorphous phase tries to separate positive and negative charges pressing them to the crystallite boundaries, however not affecting the zero field inside crystallites, as it has been already shown. Free movement of the separated charges is somewhat constrained and this can be one of the reasons why the apparent conductivity decreases during poling.

## 4. Conclusion

Some additional experimental evidence is provided on close relation between polarization and space charge in PVDF. It has been shown that the depolarizing field is compensated by charges trapped in macroscopic transition zones by which polarized parts of the volume are separated from non-polarized ones. The compensating charges are either injected in the bulk, or created inside due to the thermal emission and then trapped when the ferroelectric polarization in crystallites is formed. The clarification was necessary, because it is still believed sometimes that the poling field is created by the trapped charges [11]. It follows from our results, however, that the charge trapping in PVDF and probably in other ferroelectric polymers is not the cause, but rather the result of the polarization buildup.